\documentstyle[prl,aps,twocolumn,epsf,epsfig]{revtex}   
\newcommand{\be}[1]{\begin{equation}\label{#1}}
\newcommand{\ee}{\end{equation}}     
\newcommand{\bea}{\begin{eqnarray}}
\newcommand{\eea}{\end{eqnarray}} 
\newcommand{\eq}[1]{Eq.~(\ref{#1})}
\newcommand{\pic}[2]{\epsfxsize #1 cm
\epsffile{#2.eps}
}
\newcommand{\fig}[1]{Fig.~\ref{#1}}
\begin{document}  
\tighten   
%
\title{ \large\bf Time dependent energy absorption of  atomic clusters from an 
intense laser pulse}
\author{Christian Siedschlag and Jan M. Rost} 
\address{
 Max-Planck-Institute for the Physics of Complex Systems, 
N\"othnitzer Str. 38, D-01187 Dresden, Germany\\[3mm]
\rm (November 2000)}
\author{
\begin{minipage}{152mm}  
\vspace*{3mm}
For the energy absorption of atomic clusters as a 
function of the laser pulse duration we find a similar behavior
as it has been observed for metallic clusters [K\"oller et al., Phys. 
Rev. Lett. {\bf 82}, 3783 (1999)]. 
In both situations there exists
an optimum radius $R_{o}$ of the cluster for
energy absorption. In the metallic case the existence of 
$R_{o}$ has been interpreted as a consequence of the collective oscillation
of a delocalized electron cloud in resonance with the laser 
frequency. Here, we give evidence that in the atomic cluster the origin of 
$R_{o}$ is very different. Based on field assisted tunneling it 
can be related to the phenomenon of
enhanced ionization as it occurs in small molecules. 
The dependence of $R_{o}$ on the laser frequency 
turns out to be the key quantity to distinguish the processes.
\draft\pacs{PACS numbers:  36.40 -c, 33.80 Rv, 36.40 Gk}
\end{minipage}
}
\maketitle
 Exposed to an intense laser pulse a cluster absorbs a considerable amount
  of energy 
 which is released subsequently through 
fragmentation into fast electrons \cite{Shaal96}, multicharged ions 
\cite{Dital97} and
radiation in the x-ray regime \cite{McPal94}.
These effects depend on the seize of the cluster, i.e., the number of 
constituent atoms and the pulse duration,  somewhat less on the kind of  atoms
and the wavelength of the light.

In a recent experiment K\"oller et al. measured the  intensity dependent 
energy absorption of  a cluster consisting of some ten  platinum atoms
as a function of laser pulse  duration\cite{meiwes}. 
This has been done by keeping the 
energy content (fluence) of the laser pulse constant and varying the 
pulse duration as well as the  peak intensity accordingly. 

Interestingly, the absorbed energy {\em decreases} with {\em 
increasing} laser intensity, after having reached a maximum.  
We have found the same behavior in calculations for atomic clusters
containing a similar number of atoms.  However, as we 
will see below, the mechanism responsible for the phenomenon 
is quite different from the one described in \cite{meiwes}
for the metallic  clusters.

We will show that a sensitive indicator for
the underlying mechanism is the existence of an optimal mean internuclear distance
$R_{o}$ for energy absorption which changes with the laser frequency
$\omega$ for a cluster of delocalized electrons (a metal cluster)
while it is independent of $\omega$ in the case of an atomic
cluster.  In both cases the existence of
$R_{o}$ is also the origin of the peculiarity that the energy
absorption can decrease with increasing peak intensity of the laser,
as mentioned above.

In order to discriminate between different mechanisms we had to
choose an approach which is capable of handling, at least
in principle, both situations: atomic clusters with localized 
electrons and
delocalized electrons as they are typical for metallic behavior. 
Furthermore, the numerical treatment had to be fast enough to follow
an appreciable number of particles (ions and electrons).  Clearly,
this cannot be done fully quantum mechanically, for the time being. 
Our approach is a combination of the ones described 
in \cite{RoPal97,Dit98,LaJo00,IsBl00}, i.e., essentially based 
on classical equations of motion for all ionized
charged particles under full mutual Coulomb interactions. 
As in \cite{Dit98} we have used Coulomb soft-core potentials
\be{softcore}
V_{e} (r)= (r^2+a^2)^{-1/2}\,.
\ee
We will see later,  that the choice of $a$ allows us to describe an 
atomic cluster with localized electrons ($a_{a}\sim 1$a.u.) or to simulate 
a metallic cluster with delocalized electrons ($a_{m}\gg a_{a}$).

The initial ionization of an electron bound to an atom or ion is 
described with an analytically known rate \cite{Lifschitz}, 
dependent on the instant (static) electric field
at the position of the atom/ion to be ionized. The field is created by all surrounding charges 
(ions and electrons) and the laser. In contrast to \cite{Dit98} we do 
not include additional electron impact ionization.  Its effect is 
small (see \cite{IsBl00}), moreover, its implementation based on 
empirical cross sections, such as the Lotz formula, 
bares the danger that the contribution of 
electrons to ionization is counted twice: through field ionization 
and through additional impact ionization.

The actual computation goes as follows: After a relaxation to an 
equilibrium under Lennard-Jones potentials the atomic configuration 
is exposed to the laser pulse. We compute a probability for 
ionization for each atom (later ion)  from the rate in an time 
interval $\Delta t$. Is it larger than a generated random number $0\le 
s\le 1$, the atom is considered as 'ionized' \cite{remark} and turns into an ion, and 
a new electron is created outside the instant potential barrier with  
zero kinetic energy.  The ionization rate for the ion is adjusted to 
the corresponding higher binding energy and the procedure is repeated,
of course, simultaneously for all atoms/ions.
Newton's equations  are solved for the time 
evolution of all charged particles interacting through 
mutual Coulomb soft-core potentials \eq{softcore} 
with $a^2=2$a.u..

\begin{figure}
\begin{center}
\pic{7}{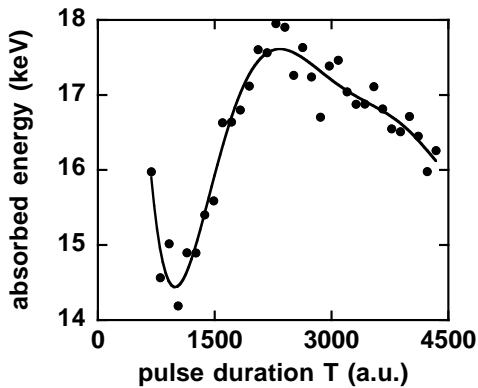}
\end{center}
\caption{Absorbed energy as a function of laser pulse duration $T$ for 
constant fluence such that a peak intensity of 10$^{15}$W/cm$^2$ is reached
with a pulse of 20 cycles ($\omega=0.055$ a.u.). Note that an atomic time unit is 0.0242 fs. The line is to guide the eye.}
\label{f:engain}
\end{figure}

In the following we will discuss the energy transfer to a Ne$_{16}$ 
cluster from a laser pulse with $\sin^2$-envelope and an optical 
frequency of $\omega = 0.055$ a.u..  If we record the  energy 
gain after the pulse as a function of pulse duration $T$, we obtain 
\fig{f:engain}.  Since for constant fluence under a variation of 
$T$,  the peak intensity behaves as $I\propto 1/T$, 
one recognizes the increasing energy absorption for decreasing 
intensity. Only for very short pulses (high intensities) the trend is 
reversed indicating that in this regime the pure atomic response 
dominates cooperate cluster effects.  An analogous behavior of the 
energy absorption, including the 
rise for very short pulses, is found 
in calculations for excitation of an Na$_{9}$ cluster \cite{privcom}, and in the 
experiment  on platinum clusters (exemplified by the 
dependence of the charge states for ejected ions as a function of the 
pulse duration, see Fig. ~2 in \cite{meiwes}).

The authors of \cite{meiwes} provided an appealing interpretation in terms
of a plasma model for the delocalized electron density of the platinum 
cluster: The eigenfrequency  $\Omega$ of the electron cloud depends on its 
density, which, in turn, is a function of the cluster 
radius, i.e. $\Omega = \Omega(R(t))$. When the 
cluster expands due to the net positive charge after initial 
ionization, the electron density decreases and so does the plasma 
frequency $\Omega$ which will eventually match the laser frequency
$\omega$. Then, energy absorption becomes resonant and 
is greatly enhanced.  The maximum in the  absorption as a function of 
pulse duration is now essentially a matching problem: The best 
condition is a coincidence of the peak intensity with the time when 
$\Omega(R(t)) = \omega$.  If the laser pulse is too short, the resonance 
condition is reached when the pulse is already over. On the other 
hand, if the pulse is too long, the cluster has expanded beyond 
$R_{o}$ when the peak intensity is reached.

We define as a characteristic length scale for the cluster
\be{MID}
R(t) = \left(\sum_{i}^N R_{i}^2/N\right)^{1/2}\,,
\ee
the mean over 
all individual internuclear distances $R_i$. Equivalently,
we will speak  of the cluster radius which 
is directly proportional to $R$ for a fixed number of 
atoms in the cluster \cite{remark}.
Typically,
$R(t)$ increases 
adiabatically slowly compared to the electronic and optical time scales.
This allows
us to gain more insight into the dynamics by considering the energy
absorption of the cluster for different but
 {\em fixed} mean internuclear separations $R$.  Figure
\ref{f:fixedR}a demonstrates that the energy absorption for fixed $R$
peaks at a critical $R_{o}$ independent of the laser frequency.

This is a key observation which has several consequences:
Firstly, the existence of $R_o$ for an atomic cluster
explains the shape of the energy absorption in \fig{f:engain} with a
maximum due to the  monotonic increase of $R$
as a function of time. Large energy absorption occurs for a
pulse duration $T$ such that peak intensity  is reached at $T/2$, when
the cluster has the optimal seize $R(T/2) \approx R_{o}$.  Secondly, the
mechanism which leads to the existence of $R_{o}$ must be different
from the one proposed for a metal cluster in \cite{meiwes}, since 
a resonant absorption with $\Omega(R_{o}) = \omega$ points to
optimal cluster radii $R_{o}$ which change with the frequency
$\omega$. Rather, the mechanism we have identified for these relatively small
 atomic clusters is akin to a behavior in small molecules which has
 been described under the name enhanced ionization \cite{Seial95} or
 CREI (charge resonant enhanced ionization) \cite{ZuBa95}.  In short,
 the idea is that in a diatomic molecule the electron, localized on one
 atom most of the time due to the oscillatory light field, can easier
 tunnel through the barrier formed by the attractive Coulomb potential
 and the electric field since this barrier is lowered by the additional
 electric field generated by the neighboring (positively charged)
 nucleus.  
	
An optimal internuclear distance $R_{o}$ for this field
 assisted tunneling exists since in the united atom limit $R=0$ there is only
 one well (and deeper binding) while in the separated 
 atom limit $R\to\infty$ the additional field simply goes to zero. The signature 
of this mechanism is the existence of $R_o$ 
and its independence of the laser frequency $\omega$, precisely 
as seen in \fig{f:fixedR}a.
Hence, the mechanism of field assisted tunneling
is indeed also operative for our cluster where many surrounding
charged ions form a strong field for the specific atom or ion to be
ionized in the cluster.
 
 Having established the origin of the peculiar behavior of energy 
absorption in an atomic cluster as a function of laser pulse 
duration,  we  may subject our modelling of  cluster dynamics to 
an ultimate test by comparing  exact 
quantum results for the simplest 
system $H_{2}^+$ to predictions from our approach.  This is done 
in one dimension 	
 \twocolumn[
\begin{minipage}{17.5cm}
\begin{figure}
\hspace*{-.6cm}\pic{6.2}{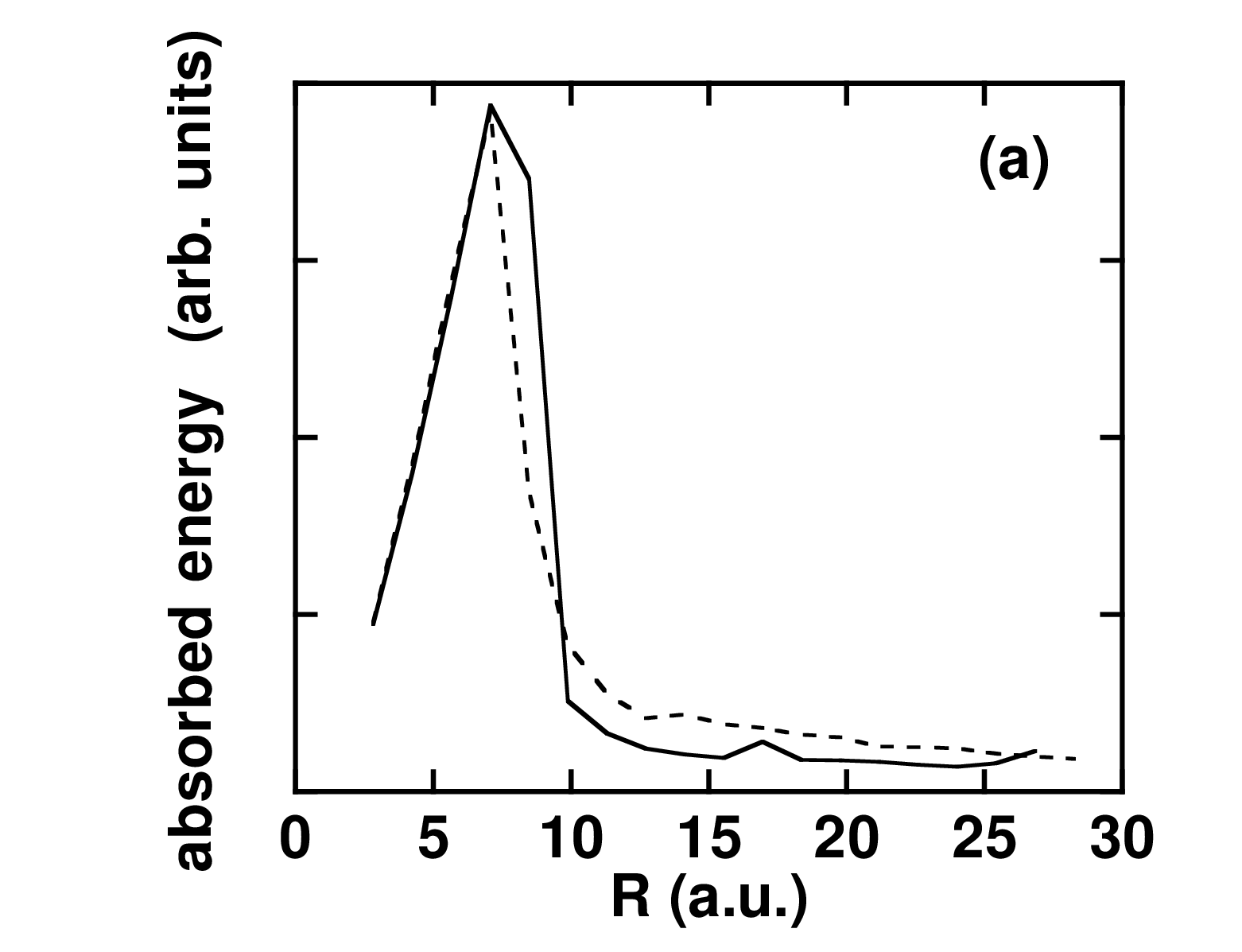}\hspace*{-.5cm}\pic{6.2}{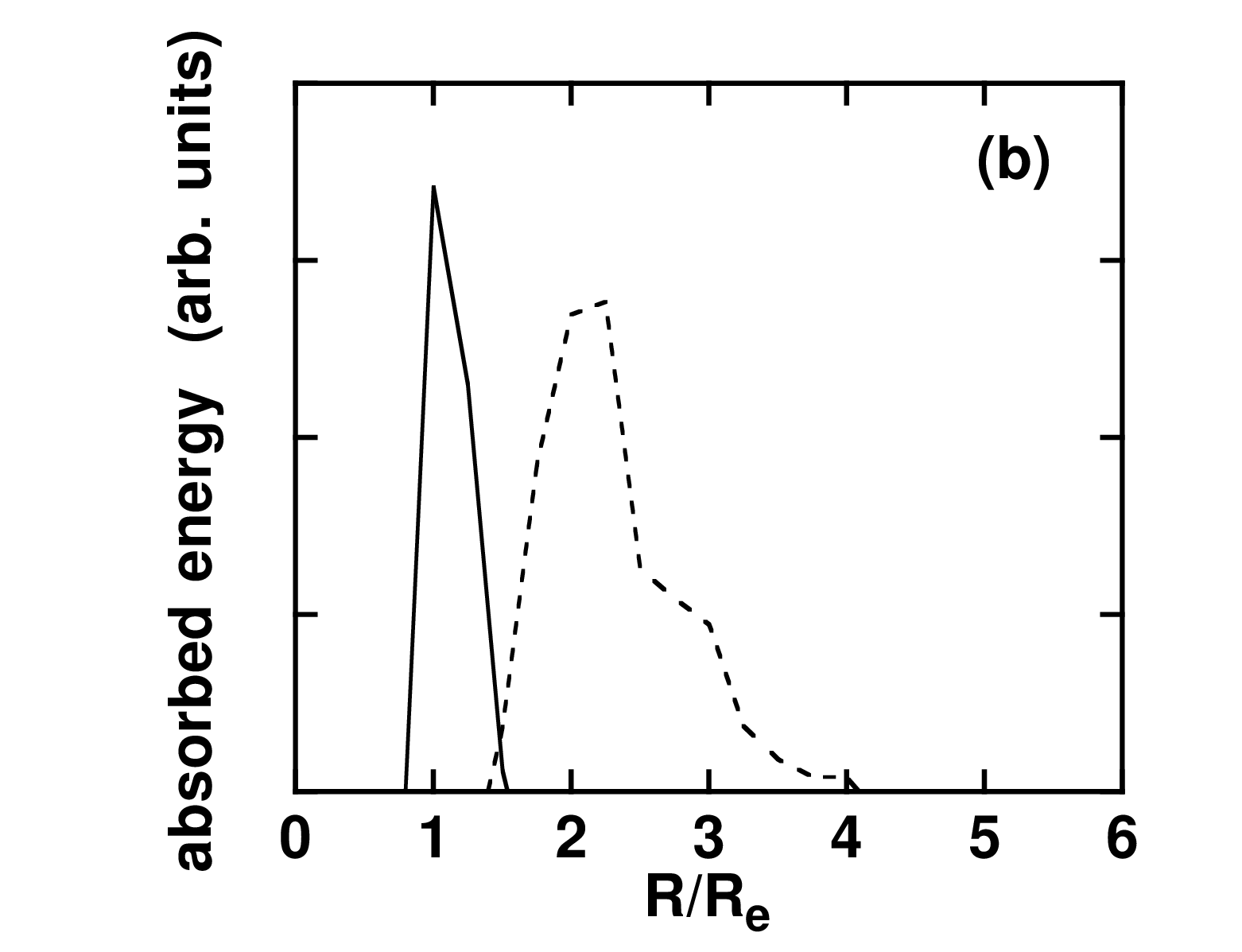}
\hspace*{-.5cm}\pic{6.2}{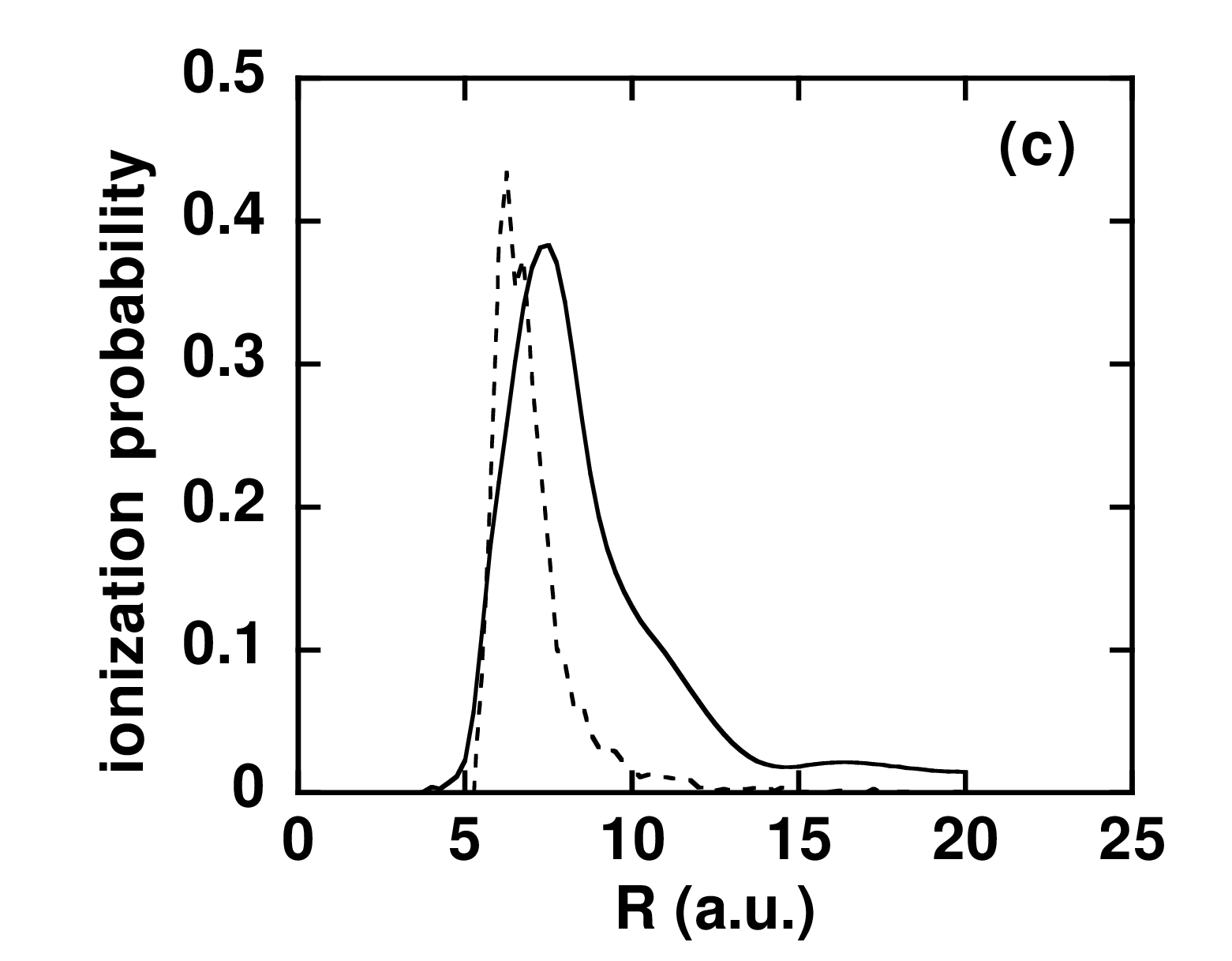}
\vspace*{3mm}
\caption{Energy absorption from an intense laser pulse ($T=55$ fs) in different 
situations: (a) for Ne$_{16}$ as a function  of fixed mean interatomic distance $R$ 
at  two different laser 
frequencies,  $\omega = 0.055$ a.u. (solid), $\omega = 0.11$ a.u.
(dashed) and with peak intensity $I = 10^{15}W/$cm$^2$ , 
(b) as in (a) but for the 16 atom metallic cluster model as a function of 
the initial mean ion distance $R_e$ and for $I = 3.51 \times 10^{12}{\rm W}/{\rm cm^2}$, see 
text, (c) for  $H_{2}^+$ as a function  of fixed internuclear 
distance  with the one dimensional quantum 
result (solid) and the present tunneling approach
(dashed) at a peak intensity of $I=5.6 \times 10^{13}{\rm W}/{\rm cm^2}$.
\vspace*{6mm}}
\label{f:fixedR}
\end{figure}
\end{minipage}
]
\noindent 
(where the internuclear axis is aligned  along the 
electric field of the laser) in \fig{f:fixedR}c.  Although we model the 
bound electron being attached to one proton and calculate its tunneling 
rate subject to the laser field and the field generated by the second 
proton, the actual ionization probability is in surprisingly good agreement 
with the exact quantum result, particularly compared to a purely classical 
over barrier model whose ionization yield is too small to be visible in  
\fig{f:fixedR}c. Note that $R_o$ for the cluster  (\fig{f:fixedR}a) is 
even roughly equal to $R_o$ in $H_2^+$ (\fig{f:fixedR}c).

Designed for an interaction of several ions with many electrons and an
intense laser pulse the fairly accurate description of $H_{2}^+$ is an
unexpected confirmation of the modelling.  However, it raises also the
question if the mechanism we have identified for energy absorption in
clusters being akin to that for molecules in intense laser fields is
merely a consequence of the modelling which seems to be ideally suited
to describe tunneling related phenomena.

\begin{figure}[b]
\pic{8}{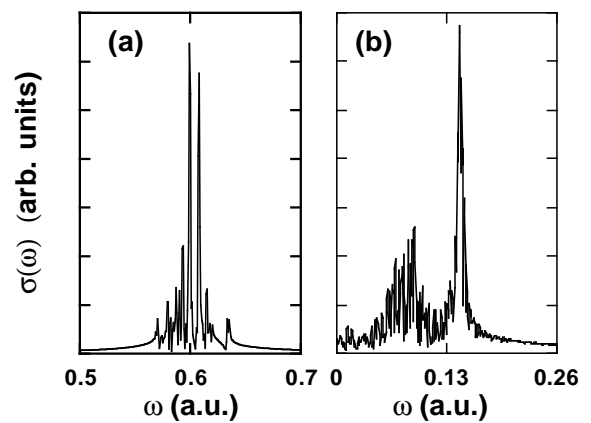}
\caption{Excitation spectrum for the electrons  in the cluster, (a) 
localized electrons in  Ne$_{16}$, (b) delocalized electrons in the 
metallic cluster model.}
\label{f:spec}
\end{figure}

To doublecheck that our result is independent of the modelling and also, 
to clarify further the  different mechanism  which seems to be 
responsible for the (similar) energy absorption and existence of a 
critical mean distance $R_{o}$ in metal clusters we have simulated {\em within our 
approach} the behavior of delocalized electrons as they occur in a 
metal cluster.  This has been achieved by artificially softening the 
potential \eq{softcore} with  $a^2=30$~a.u..
As a consequence the cluster ions at equilibrium distance of each 
other form one structureless well for the "collective " binding of 
the electrons.  Comparing the excitation spectrum  of the electrons,
one sees for the original situation of atomic clusters with localized 
electrons a single peak which corresponds to the local excitation
(\fig{f:spec}a) while for the delocalized electrons  with $a^2=30$ one 
sees two peaks (\fig{f:spec}b), the lower and wider one corresponds to the softened 
local  excitation of the local binding, the higher peak is the new 
feature of collective excitation which is believed to be responsible 
for the mechanism of resonant energy absorption as described above. 
If this is true, we would expect in our model for delocalized 
electrons  a  dependence of the optimal cluster radius for energy 
absorption on the laser frequency. This is indeed the case, as one  
can see in \fig{f:fixedR}b.  As expected for  a 
decreasing electron density with growing cluster radius, and 
corresponding decreasing eigenfrequency $\Omega$ of the electron 
cloud, we find that $R_{o}$ is smaller for the higher laser frequency.

This confirms  the existence of a different mechanism which leads to a 
critical cluster radius in a situation of delocalized electrons, in 
accordance with what has been found by very different modelling 
of the valence electrons in sodium clusters \cite{PG1,PG2}. It
also demonstrates that  our 

\noindent
formulation of intense field dynamics of 
clusters is capable of  describing both,  atomic clusters with 
localized electrons and, at least qualitatively, the situation of delocalized electrons 
as they occur in metal clusters.   Hence, 
the result  reassures that 
the mechanism of field enhanced ionization by surrounding charged 
particles is not an artifact of the  theoretical description.

To summarize, we have found that the energy 
absorption in small atomic clusters depends strongly on the
laser pulse duration, similarly as in metallic clusters 
and large ($N\approx 10^6$) atomic clusters. However,
the mechanism is very different. While in metallic cluster 
\cite{meiwes} as well as in large atomic clusters \cite{ZDP99} 
a similar plasmon resonance mechanism prominently involving  delocalized electrons
has been advocated to explain the
observations, we find that in small atomic clusters field assisted
tunneling is responsible.
By making use of the adiabaticity of the ionic
motion compared to the electronic motion we could show that a critical
cluster radius exists for maximum energy absorption which is
{\em independent} of the laser frequency $\omega$.  This behavior is akin to
the one known from diatomic molecules as ''enhanced ionization
"\cite{Seial95,ZuBa95} and can be attributed to the same physical
effect of field assisted tunneling ionization.  As a sideffect, we
have shown that our approach also describes the ionization of the
smallest molecule, $H_{2}^+$, in a strong laser pulse rather well.

Furthermore, we have  simulated the behavior of 
delocalized electrons within the same theoretical approach. 
Thereby, we could confirm
that for delocalized electrons enhanced energy absorption can be 
attributed to a plasmon type resonance. It occurs when
 the eigenfrequency of the delocalized 
electron density and the laser frequency agree, as suggested  by K\"oller etal to 
interpret their experiment \cite{meiwes}. However, we could only clearly 
identify this type of resonance behavior if exclusively the valence electrons 
are involved in the ionization dynamics, i.e., if the laser intensity 
is sufficiently weak (in our case $3.51\times10^{12}$ W/cm$^2$). Once electrons from the 
ionic cores are ionized, the local character of the electron binding 
starts to dominate. Moreover, 
field ionization triggered by surrounding charges takes over the 
ionization caused by the collective electron cloud and the laser 
field. Since in the experiment \cite{meiwes}  the peak intensity of the 
laser was rather large (more than $10^{15}$ W/cm$^2$) and highly charged ions  have 
been detected (which probably were even higher charged through the 
initial ionization before recombination took place),  it is possible
that the actual mechanism for the energy absorption 
pattern as a function of pulse duration is closer to that of 
field assisted tunneling  as in atomic clusters than to the plasmon resonance
enhanced  ionization of metallic valence electrons.
As we have shown, the two 
mechanisms differ by their dependence on the laser frequency. Hence, 
it would be desirable to repeat the experiment of \cite{meiwes} at a higher laser 
frequency, the best choice being an $\omega$  
high enough that the resonance condition cannot be 
fullfilled. If the energy absorption pattern still shows a pronounced 
maximum, one could exclude the plasmon induced absorption mechanism 
and rather would  have to conclude that the field assisted tunneling 
scenario is a universal mechanism for intense laser field ionization 
in molecules and clusters of moderate seize.

It is a pleasure to thank   K.H. Meiwes-de Broer, P. Corkum, R. Schmidt, 
and P.-G. Reinhard 
for fruitful discussions. We also acknowledge O. Frank's input at the 
initial stages of this work  which has been supported by the DFG 
through the Gerhard Hess-program.


\begin{references}
\bibitem{remark} 'Ionization' means here that the electron is no 
longer bound to its mother atom, however, it may be still bound to the 
entire cluster. See also the distinction of 'outer' and 'inner' ionization 
in \cite{RoPal97}.
\end{references}
\end{document}